# Automatically Growing Global Reactive Neural Network Potential Energy Surfaces: A Trajectory Free Active Learning Strategy

Qidong Lin[1], Yaolong Zhang[1], Bin Zhao[2], and Bin Jiang[1,*]

[1]*Hefei National Laboratory for Physical Science at the Microscale, Department of Chemical Physics, Key Laboratory of Surface and Interface Chemistry and Energy Catalysis of Anhui Higher Education Institutes, University of Science and Technology of China, Hefei, Anhui 230026, China*
[2]*Theoretische Chemie, Fakultät für Chemie, Universität Bielefeld, Universitätsstraße 25, D-33615 Bielefeld, Germany*

*: corresponding author: bjiangch@ustc.edu.cn

**Abstract**

An efficient and trajectory-free active learning method is proposed to automatically sample data points for constructing globally accurate reactive potential energy surfaces (PESs) using neural networks (NNs). Although NNs do not provide the predictive variance as the Gaussian process regression does, we can alternatively minimize the negative of the squared difference surface (NSDS) given by two different NN models to actively locate the point where the PES is least confident. A batch of points in the minima of this NSDS can be iteratively added into the training set to improve the PES. The configuration space is gradually and globally covered with no need to run classical trajectory (or equivalently molecular dynamics) simulations. Through refitting the available analytical PESs of $H_3$ and $OH_3$ reactive systems, we demonstrate the efficiency and robustness of this new strategy, which enables fast convergence of the reactive PESs with respect to the number of points in terms of quantum scattering probabilities.

Keywords, potential energy surface, data sampling, neural networks, active learning

## I. INTRODUCTION

As a natural consequence of the well-known Born Oppenheimer approximation, potential energy surface (PES) plays a central role in chemical physics, serving as an initial and necessary ingredient for extensive spectroscopic and dynamical simulations. The modern way of constructing an accurate PES requires at least two key steps, namely the ab initio calculations at scattered molecular configurations and the mathematical form linking the relationship between molecular configurations and potential energies. As the complexity of the system increases, it becomes increasingly difficult to sample data points in the configuration space and to find empirical potential energy functions with sufficiently high accuracy[1], especially for reactive systems. Instead, it is desirable to find an automatic way to construct high-dimensional PESs using a black-box like fitting (or interpolation) method and data selection scheme. Several pioneering fitting/interpolation methods have been developed for small/medium systems in gas phase since 1990s, such as modified-Shepard interpolation (MSI)[2,3], interpolating moving least-squares (IMLS)[4,5], permutationally invariant polynomial (PIP) fitting[6,7], and so on. More recently, the emergence of the powerful machine learning (ML) methods, *e.g.* NNs and Gaussian process regression (GPR), have enabled more accurate and scalable PESs from gas phase reactive systems[8-12], to gas-surface reactions[13-20], and to materials[21-23].

On the other hand, the choice of data points to perform ab initio points is equally important when constructing an accurate PES. In general, for a given reactive system, one requires the PES to be globally accurate in the region where the quantum/classical reaction dynamics simulations would access. In this respect, a wise data selection scheme would typically need much fewer points than random or regular grid sampling, while yielding a PES with comparable or even higher level of accuracy. Collins and coworkers[2, 3] have pioneered the use of classical trajectory (or equivalently molecular dynamics) calculations in the MSI method to efficiently sample the dynamically important regions in the configuration space and iteratively add new points where the trajectories visit most frequently and where the PES is most inaccurate. This strategy has later been widely transferred to other fitting approaches with different modifications or using direct dynamics trajectories[8, 10, 15, 18-20, 24-34]. For growing NN-based PESs, for example, Raff et al. have proposed a probabilistic criterion to select new configurations in a region with a sparse density of points[24]. Behler has suggested to identify the poorly sampled region of the configuration space by comparing the deviation of the predicted energies by different NN fits[25]. Zhang and Guo groups have taken both advantages adding a new point with a sufficiently small generalized Euclidian distance with existing points and with sufficiently large energy deviation among several NN fits[8, 10, 26-28]. Similarly, Ceriotti et al.[29] and Dral et al.[30] proposed the structure-based farthest point sampling scheme tending to select points far away from others.

These trajectory-based iterative sampling algorithms can be actually classified as "active learning" in the language of machine learning, which is a process of using previously learned information to actively drive the augmentation of the training data. Active learning strategies have also been applied in deep NN[35] and GPR[36-39] methods.

Interestingly, the GPR model not only predicts the potential energy of a given geometry but also provides the variance of the prediction. This unique feature allows one to find the maximum in the variance space where the prediction is least confident, which naturally becomes an appropriate location to include a new point, as done by Guan et al.[37] and Toyoura et al.[36] Very recently, Vargas-Hernández *et al*. further proposed a Bayesian optimization based algorithm to identify new points that mostly improve the calculated quantum scattering observables[39]. A common feature of these GPR-based active learning procedures is that they need no classical trajectories to explore the configuration space, which could be potentially more efficient and more ergodic than the trajectory-based sampling. This is particularly true for reactive systems with high barriers and/or deep wells, where many trajectories would repeatedly access the same area in practice.

Apparently, it is not straightforward to combine such a variance searching algorithm with other fitting or interpolation methods which provide no information on the uncertainty of the prediction. Interestingly, Uteva et al. have shown that more accurate PESs with fewer points can be obtained by selecting points based on the largest discrepancy between two different GPR models rather than the highest variance of a single GPR model[38]. This motivates us to propose here a trajectory-free active learning algorithm applicable to NN PESs. Instead of minimizing the variance in a GPR model, we add a batch of new points at local minima of the negative of the squared difference surface (NSDS) given by two (or more) independent NN fits multiplied by an energy-dependent weighting function. As pointed out by Dawes *et* *al.*[5] in the study of the IMLS method, the NSDS, which is supposed to approach zero at existing data locations while has some minima in between, is essentially a measure of the uncertainty of prediction. We observe the same phenomenon using NNs, as shown below.

In the following, Sec. II describes the data selection procedure in detail based on searching the local minima on the NSDS. In Sec. III, we test this new scheme in two benchmark small reactive systems, namely H + $H_2$ and H + $H_2O$ reactions. It is validated to generate accurate reactive NN PESs for these two systems with much fewer points than previous trajectory-based methods. We finally conclude in Sec. IV.

## II. METHOD

It is well-known that NN-based potentials, or actually any PESs represented by "non-physical" functions, have very limited extrapolation capabilities[25]. Consequently, different NN potentials can give similar predictions at or nearby the existing data points but unpredictable values in the regions where no or very few points exist, giving rise to large prediction errors. There must exist a lot of local maxima (minima) on (the negative of) the squared difference surface between two NN fits among adjacent points. The regular NSDS can be written as,

$$D(\mathbf{x}) = -\left(y_1(\mathbf{x}) - y_2(\mathbf{x})\right)^2, \qquad (1)$$

where $y_1$, $y_2$ are the outputs of two NN models and $\mathbf{x}$ denotes the collection of all atomic coordinates. Since NNs are analytical functions so that the resultant NSDS is also smooth and continuous, it is very convenient and efficient for optimization. While any nonlinear optimization algorithm can be used, we choose the robust

Levenberg-Marquardt (LM) algorithm in this work to search the local minima of the NSDS.

We first demonstrate this concept in a simple one-dimensional (1D) system. A Morse potential of H$_2$ was used as the target potential energy function, namely,

$$V(r) = D_e \left( e^{-2a(r-r_e)} - 2e^{-a(r-r_e)} \right), \tag{2}$$

where $D_e = 4.74$ eV, $r_e = 1.40$ bohr, $a = 1.03$ bohr$^{-1}$. For simplicity, only a limited range, *i.e.* $r \in [0.68, 4.74]$ bohr was considered, covering the energy range of 0~6 eV relative to the equilibrium geometry. To begin with, we placed three seed points at the equilibrium and two endpoints, respectively, as shown in Fig. 1a. Given this small size of training set, two trial NNs containing one hidden layer with two and three neurons were trained. Due to the lack of points, the squared difference between the two NN fits (black curve in Fig. 1) is very large except at the three seed points and shapes into two prominent maxima in between. Starting from random initial guesses nearby existing points with a small deviation, we quickly located the deepest minimum on the NSDS (equivalently the left maximum in the Fig. 1a) with the LM algorithm and added the next point there. The updated NN fits are shown in Fig. 1b, where the short range potential has been much improved. Also shown is the updated curve of the squared difference on which only a large maximum remains. This procedure was repeated automatically until the NN fit was accurate enough. Specifically, three points were selected during this process (6 points in total) and the root-mean-square error (RMSE) between the final NN potential and the Morse function, estimated from 50 randomly distributed points in the energy range of our interest, was only 0.089 meV (Fig. 1d). We found that the total number of points required to reach this level of accuracy is insensitive to the trial NN fits used to define the NSDS and the initial guesses in the LM algorithm. This 1D test clearly illustrates how this data sampling scheme works.

In more complex systems, searching local minima on the regular NSDS defined by Eq. (1) can be sometimes problematic. The optimization would tend to find some very deep minima on this NSDS at those unphysical or highly-distorted configurations, especially when the data set is small. Iteratively adding these high energy points into the training set would significantly decrease the fitting accuracy in the low energy regions which are most important for dynamics simulations. To avoid this problem, we modify the NSDS by introducing an energy-dependent weighting function similar to that was used by Guan et al.,[37]

$$\omega(y_1, y_2) = \exp\left(-\alpha\left(\frac{y_1^2(\mathbf{x}) + y_2^2(\mathbf{x})}{2}\right)\right), \tag{3}$$

where $\alpha$ is an adjustable parameter. Assigning energy zero to the global minimum of the system, $\omega$ would decay exponentially as energy increases. The weighted NSDS,

$$F(\mathbf{x}) = \omega(y_1, y_2) D(\mathbf{x}), \tag{4}$$

enables the optimization favoring low energy regions, which is a better choice for our purpose. Fig. 2 illustrates the damping function $\omega(y) = \exp\left(\frac{-\alpha y^2}{2}\right)$ as a function of energy with a varying $\alpha$. One could easily correspond a specific $\alpha$ value to an energy range of interest. For example, $\omega$ decreases to almost zero at energy of 2 eV with $\alpha$=1.5 eV$^{-2}$, which will vanish the weighted NSDS above 2 eV with its gradient approaching zero. This will definitely avoid accessing high energy regions during the optimization process. As $\alpha$ decreases gradually, the accessible energy interval becomes increasingly broader so that there is an increasingly higher opportunity to select some high energy points. In other words, the value of $\alpha$ controls the energy distribution of the sampled points.

For multi-dimensional reactive systems, a pragmatic procedure of actively selecting data points is given as follows and illustrated in Fig. 3. It is desirable to start with a minimum number of seed points, *e.g.* taken from the minimum energy path (MEP) which can be nowadays routinely obtained from ab initio quantum chemistry packages, or even an intuitive guess. Since NNs are highly nonlinear functions, it is suggestive to train several trial NNs with different initial guesses and choose the best two out of them to evaluate the NSDS. Optimizations on the NSDS can be initiated from randomly sampled configurations close to but not exactly coincide with existing configurations (otherwise the gradient can be zero). Many local minima on the weighted NSDS may be found in such cases, which are sorted in a waiting list in descending order by their well depth. Some of them may be identical (*e.g.* due to symmetry) and have to be excluded. In addition, if two minima are very close to each other, *e.g.* in light of their Euclidean distance, the shallower one should be also removed from the list. For the remaining local minima, we will also calculate the absolute energy discrepancy ($\Delta E$) between the two trial NNs without multiplying the weighting function, which has a dual effect in data selection. First, if $\Delta E$ of a given point is less than certain predefined threshold ($t_1$), *e.g.* the desired accuracy of the PES, it is unnecessary to add this point. Although this could happen near an accidental crossing of the two trial NN fits, it is more likely due to the fact that the weighting function is too strongly biased to low energy region where the PES is sufficiently accurate. $\alpha$ should be decreased in the next iteration as a consequence of this fact. In this scenario, the second role of $\Delta E$ is a regulatory factor to dynamically adjust $\alpha$ so that the PES is gradually grown in the configuration space from low to high energy regions. Start with a relatively high initial value ensuring the sampling of low energy regions in the beginning, $\alpha$ needs to be stepwise lowered, if $\Delta E$ is less than another threshold ($t_2$, in general, $t_2>t_1$) at a local minimum. Finally, the surviving points that fulfill these requirements will be ultimately accepted for new ab initio calculations and a new iteration will begin with an updated data set.

In order to test this trajectory-free active data placement algorithm, we applied it to two benchmark reactive systems, namely the three-dimensional $H + H_2 \rightarrow H_2 + H$ reaction and the $H + H_2O \leftrightarrow OH + H_2$ reaction including six internal degrees of freedom. On the same global PES, the latter system also involves an exchange reaction channel, i.e. $H + H'OH \leftrightarrow HOH + H'$. Avoiding the high computational cost of performing new ab initio calculations, we instead used the available highly-accurate analytical PESs for the two systems, which is sufficient for our purpose. We adopted the permutation invariant polynomial neural network (PIP-NN) approach[10, 40, 41] to account for the crucial permutation symmetry in these two systems. Three and twenty-two permutation invariant polynomials calculated by monomial symmetrization[42] were used in the input layer of NNs. To check the convergence of the PESs, we carried out quantum reactive scattering calculations for two systems studied here. The total reactive probability of zero total angular momentum for each reaction was calculated based on a quantum wave packet approach[43]. Since the quantum reactivity depends sensitively on the global feature of the PES way beyond the MEP, such calculations provide the most stringent validation of the accuracy of the PES and the efficiency of the data sampling approach.

## III. RESULTS AND DISCUSSION

### A. H+H$_2$ Reaction

For the H+H$_2$ reaction, all potential energies were computed on the well-known BKMP2 PES[44]. To start with, twelve points were placed evenly along the MEP as seed points. Two different PIP-NN structures were applied consisting of two hidden layers with 10 and 12 neurons in each layer, labeled as 3-10-10-1 and 3-12-12-1 respectively. In each iteration, we searched local minima on the weighted NSDS using the LM algorithm initiated from 300 configurations deviating from existing points by random displacements. The distance between two optimized minima was decided by Euclidean distance in terms of internuclear distances, namely $d = \sqrt{\sum_i^N (r_i - r_i')}$, where $r_i$ corresponds the *ith* bond length and $N$ is the number of bonds. Once $d<0.1$ Å between two minima, the shallower one was discarded. In addition, the point with $\Delta E$ less than 20 meV was excluded. The resulting six deepest local minima (if there were) were added to the training set. The initial value of α was set to be 1.5, which was reduced by 0.10 eV$^{-2}$, if $\Delta E$ of one of the six points was less than 80 meV, until it reached the lower limit of 0.2 eV$^{-2}$, covering a reasonable energy range up to 5 eV.

It should be noted that the fitting errors of the PIP-NN potentials are typically very small, regardless of the iteration, given the small size of training set. This gives us useless information on the accuracy of the fit. Instead, we randomly sampled 5000 points in the energy range of [0, 5] eV from the BKMP2 PES, in a cubic box defined by $R_{H1H2} \in$ [0.8, 8] bohr, $R_{H2H3} \in$ [0.8, 8] bohr, and $\theta_{H1H2H3} \in$ [0°, 180°]. These points serve as an extensive test set to assess the fidelity of PIP-NN PES.

The test RMSE is displayed as a function of the number of training points in Fig. 4. It is seen that the test RMSE is extremely large in the beginning, indicating that the initial PES is highly inaccurate. The test RMSE decreases rapidly with the increment of training points in the first few iterations. It is not surprising to see some oscillations because training NNs themselves is a highly nonlinear process so that the RMSE is sensitive to the initial guess of NN parameters, especially in this case with the small number of data points. The test RMSE drops down more slowly as the total amount of data exceeds one hundred or so, where the intermediate PES is already reasonably accurate with the magnitude of RMSE being ~100 meV. The data selection was stopped with 184 points and a test RMSE of 14.6 meV, smaller than the threshold value $t_1$. Quantum reaction probabilities on the resulting PIP-NN and the BKMP2 PESs are compared in Fig. 5, which are almost indistinguishable. This validates the high accuracy of the yielded PES.

It is more interesting to discuss the distribution of the data points selected in this way. Fig. 6a shows the energy distribution of the 184 points, which, as expected, is more biased towards low energy components. However, the population of high energy points is nonnegligible, which is necessary to eliminate unphysical holes at certain heavily-distorted configurations and warrant the convergence of quantum wavepacket calculations. Fig. 6b displays the space distribution of the 184 points projected on the contour plot as a function of the breaking and forming H-H distance and the enclosed angle optimized. Apparently, these data points are more concentrated in the interaction region and sparsely scattered in the reaction and product channels. This nonuniform distribution covers the global configuration space in an efficient way, which is a desired advantage of our sampling scheme. It is because that local minima on the weighted NSDS are more likely present where the energy varies strongly, *e.g.* the interaction region, which guarantees an adequate description of the reaction process.

**B. H+$H_2$O Reaction**

We further test our sampling scheme in the 6D $OH_3$ reactive system based on the CXZ PES reported by Zhang and coworkers[8], which itself is an accurate NN potential fitted to high level ab initio calculations. The energy zero was placed in the reactant asymptote of the H + $H_2$O channel. Again, we started with 22 seed points uniformly selected from the MEP of the abstraction reaction H + $H_2$O ↔ $H_2$ + OH. Two PIP-NN models were trained, whose architectures can be labeled as the 22-20-80-1 and 22-22-82-1, respectively. Other parameters used to control the active data selection process were kept the same as that used in the $H_3$ case, except that the decreasing rate of α was 0.01 eV$^{-2}$, the point with Δ$E$ less than 40 meV was excluded, at most twelve deepest local minima were accepted as new data points in each iteration.

Approximately 40,000 points were sampled randomly from the CXZ PES in the region of each OH bond within [0.8, 15] bohr and the energy range of [0, 5] eV, again serving as a test set to check the accuracy of the PIP-NN PES. Similarly, Fig. 7 shows that the test error decreases with the number of points, very quickly in the beginning and more gently as more data points (*e.g.* ~600) were included. The augmentation of the data set was terminated with 3067 points and the testing RMSE of the resulting PIP-NN PES was 36.8 meV (the fitting RMSE was 3.2 meV). Fig. 8 compares several representative potential energy curves of the PIP-NN and the CXZ PESs, as a function of internal coordinates deviating from the transition state of the abstraction reaction. The almost invisible differences between the two PESs indicate the high accuracy of the PIP-NN PES.

The global accuracy of the PIP-NN PES is further validated by quantum reaction probability calculations. The results for the H + $H_2O$ → $H_2$ + OH abstraction reaction and its reverse reaction, as well as the H + H′OH ↔ HOH + H′ exchange reaction are all shown in Fig. 9. It is found that the reaction probabilities on the PIP-NN PES agree extremely well with that on the original CXZ PES over the entire energy range for each reaction. This is indicative that the global configuration space has been well explored during the optimization on the NSDS. It is very encouraging since these seed points account for the abstraction reaction channels only. In this respect, the trajectory-free algorithm could be potentially superior to the trajectory-based counterpart in sampling the low probability channel.

Fig. 10a shows the energy distribution of the selected points, which is somewhat different from that for the H + $H_2$ reaction. These low energy points <1 eV largely correspond to the H + $H_2O$ and OH + $H_2$ asymptotic channels, where are well described by just a small amount of points. Many more points are focused on the 1~2 eV, corresponding to the interaction region of both abstraction and exchange reactions. Indeed, the barriers for the H + $H_2O$ → $H_2$ + OH abstraction reaction and for the H + H′OH↔HOH + H′ reaction are 0.928 and 0.892 eV on the PIP-NN PES, respectively. These values are in excellent agreement with those on the original CXZ PES. Interestingly, the nonuniform behavior is manifested in the space distribution as well, which is depicted in Fig. 10b as a function of the two longer O-H bond lengths. It is clear that the interaction region, where both O-H bond lengths are relatively short, is filled with most points. In turn, the data points in the reactant and product valleys are rather sparse. Moreover, this space distribution looks very similar to that in the work of Zhang and coworkers (Fig. 1 in Ref. [8]), except with a much lower data density. This implies that both our trajectory-free and their trajectory-based sampling algorithms have the ability and prefer to include molecular configurations in the dynamically important regions, which is most essential to accurately describe the reaction process. A key advantage of our method is that it could sample much fewer points to yield the PES at the same level of accuracy, as discussed below.

## C. Discussion

These aforementioned tests validate the high efficiency of our trajectory-free sampling scheme. It is interesting to check the robustness of this sampling scheme with respect to several parameters. To this end, we perform different data selection processes for the H + $H_2$ reaction from the same initial data set, by adjusting the weighting factor α, the decay rate of α, as well as the number of new points accepted in each iteration. These results are compared in Fig. 11 and Fig. 12. Although these different setups change the data augmentation process and the exact locations of the chosen points, they do not alter the overall space distribution of the data points, as shown in Fig. 11. The dynamically important region can be largely covered regardless of the parameters. Note that, due to the nonlinearity of NNs, even two sampling processes initiated from the identical data set with identical parameters would not eventually generate exactly the same final data set. In addition, the test RMSEs decrease similarly with regard to the number of points using various sets of parameters and they reach the same level of accuracy (~20 meV) with about $200 \pm 20$ points (see Fig. 12a). These resulting PIP-NN PESs are all found to give very converged quantum reaction probabilities (see Fig. 12b), which further validates the robustness of our sampling scheme. We note in passing that the number of random initial guesses in the LM algorithm has even a negligible influence on the data sampling process, so the results are not shown.

It is worthwhile to compare the proposed new sampling scheme with other counterparts in terms of the efficiency of data selection. For the H + $H_2$ reaction, Jiang and Guo[40] carefully sampled ~1000 data points in separate channels where the discrepancy between the fitted and BKMP2 PESs is large. High energy points and the new point with a large Euclidean distance to existing points were also discarded. For the OH$_3$ reactive system, when constructing the original CXZ PES, Zhang and coworkers[8] have collected more than 16 thousand points, using an elegant trajectory-based iterative sampling method with both the distance and energy criteria. In both cases, our sampling scheme generates less than one fifth of that amount of data to yield the equally accurate NN PESs. Moreover, Guo and coworkers[41, 45] have used ~31 and ~57 thousand points, sampled by a similar trajectory-based iterative procedure, to reproduce the CXZ PES with the PIP-NN and atomistic NN methods, respectively. Indeed, recently reported ab initio NN PESs for several representative triatomic[46, 47] and tetratomic[48-51] reactions using sophisticated trajectory-based sampling methods, have typically included thousands and tens of thousands of data points, respectively. In this regard, the present trajectory-free sampling scheme is anticipated to substantially reduce the size of data set for obtaining globally accurate reactive NN PESs.

It should be noted the GPR method is able to generate a comparably accurate PES with fewer data points than NNs[11, 12, 37, 39]. However, training and evaluating the GPR-based PES become exponentially more expensive with the increasing number of points involved. In comparison, once trained, the efficiency of the NN-based PES depends on the number of parameters only and the size of training set no longer matters. We note, although not being done in this work, that the atomic forces can be also trained by NNs when available, together with the energies. This can significantly reduce the amount of data points needed[20]. It is also interesting to test how this method will work to construct a realistic ab initio PES where no prior knowledge of the system. In such a case, having a precalculated test set with thousands of points is no longer possible. One may instead need to converge the specific dynamical property itself or the reaching a stable and small training RMSE, with respect to the increasing number of points. More tests in these aspects are desirable in the future.

## IV.    CONCLUSION

To summarize, we propose in this work an active learning strategy to iteratively add new data points in constructing NN potentials. We define the negative of squared difference surface between two different NN fits, serving as the measure of uncertainty of the NN potentials. This is an effective remedy for NNs which can not provide the variance of prediction, like the Gaussian process regression method. We can therefore add the new points at the local minima on this surface where the PES is most likely unreliable. This is repeated until the PES is sufficiently accurate. A key feature of this strategy is that it is self-driven to explore in the configuration space, with no need to run classical trajectory. This new data sampling scheme is validated to be able to generate training data with the physically correct energy and space distributions, in two prototypical reactive systems. Specifically, as few as 184 and 3067 data points are sufficient to cover the dynamically important regions and yield very accurate NN PESs for $H_3$ and $OH_3$ reactive systems, confirmed by quantum reaction probability calculations. The present trajectory-free algorithm seems to be very promising as an efficient alternative to the conventional trajectory-based counterparts for constructing globally accurate reactive NN PESs with a minimal amount of data.

**Acknowledgements**: This work was supported by National Key R&D Program of China (2017YFA0303500), National Natural Science Foundation of China (21722306, 91645202), and Anhui Initiative in Quantum Information Technologies (AHY090200).

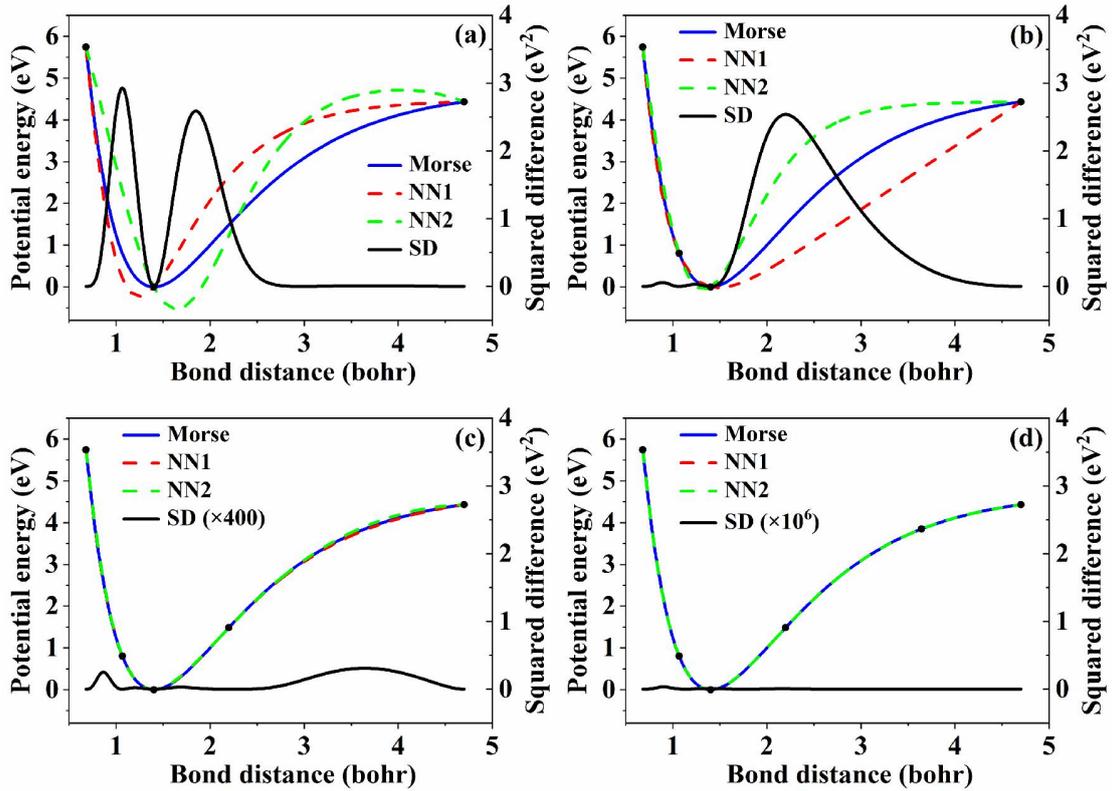

Fig. 1. One-dimensional illustrative example of the trajectory-free active learning strategy. Two independent NN models including one hidden layer with two (red dotted curve) and three (green dotted curve) neurons, respectively, were used to fit the Morse potential (blue solid curve) of $H_2$. The squared difference curve (black solid curve) is also shown for identifying the maximum of discrepancy between the two NN fits, where a new point is added. Panels (a)-(d) show the process of iteratively adding one new point to the global maximum of the squared difference curve starting from three seed points.

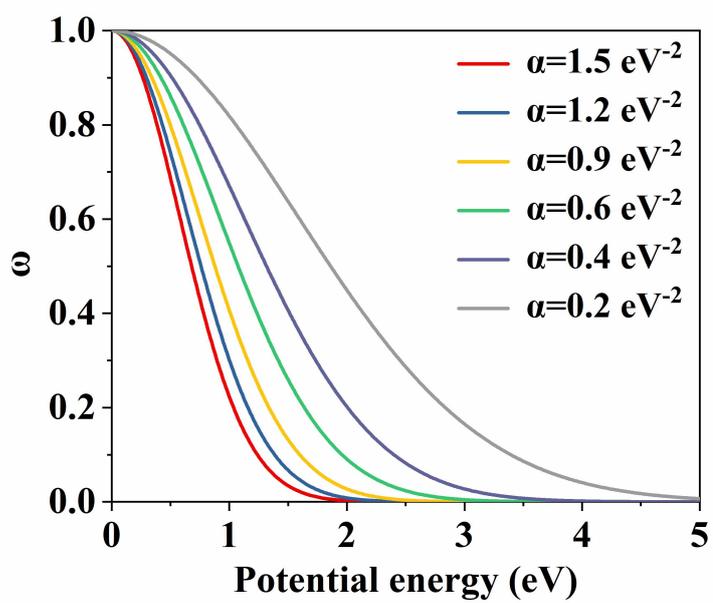

Fig. 2. The weighting function $\omega(y)$ as a function of energy with a varying α.

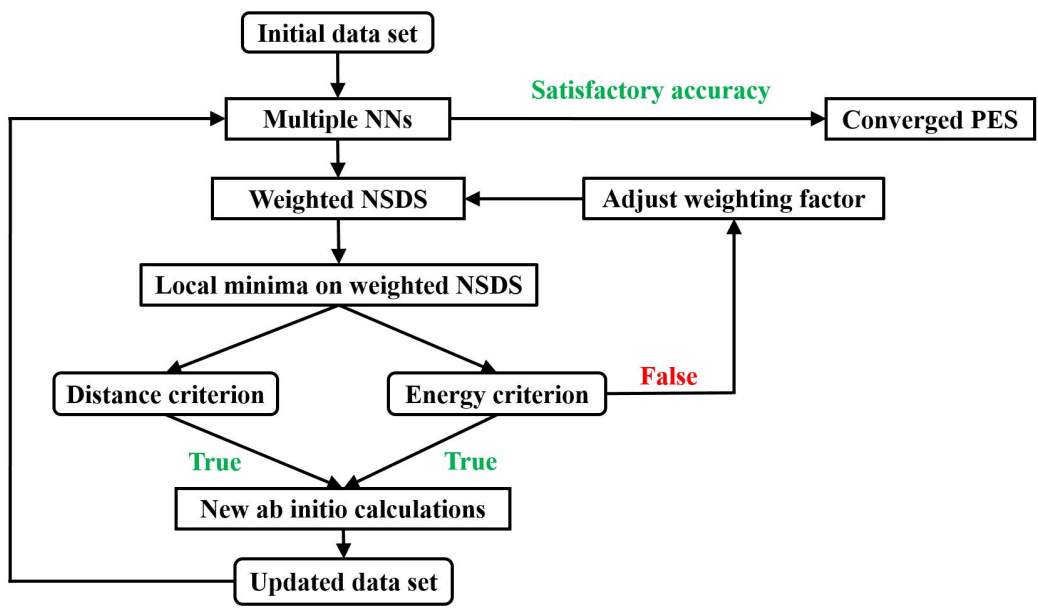

Fig. 3.The pragmatic procedure of the proposed data sampling algorithm for multi-dimensional reactive systems. NSDS stands for the negative of the squared difference surface between two NNs.

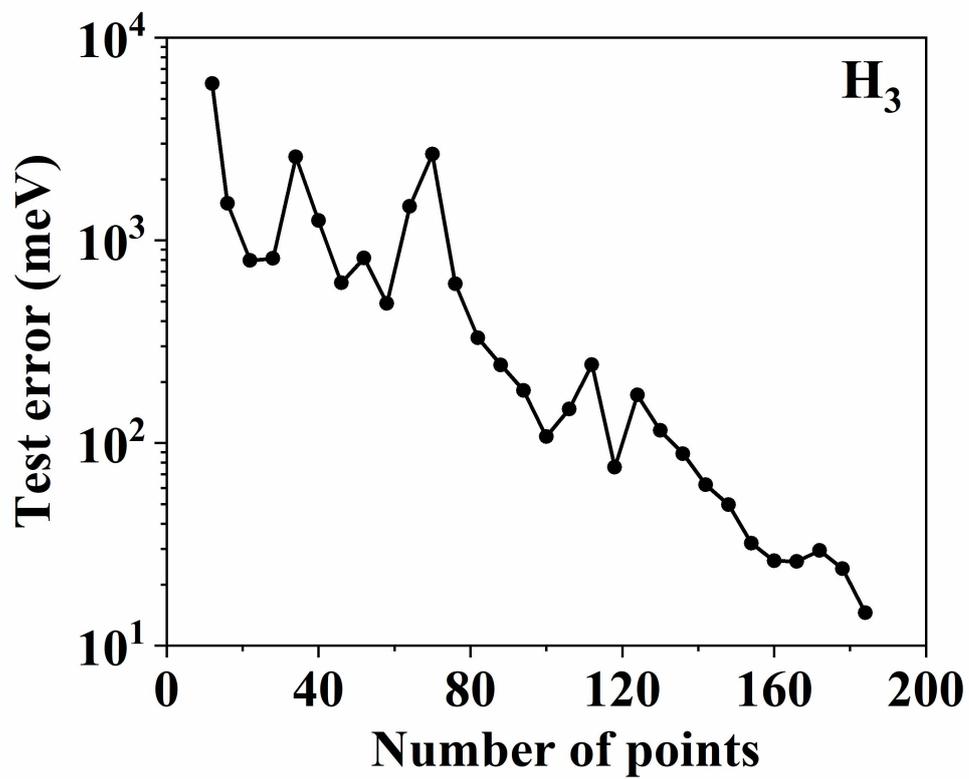

Fig. 4. Test root-mean-squared-error as a function of the number of training data points for $H_3$ system.

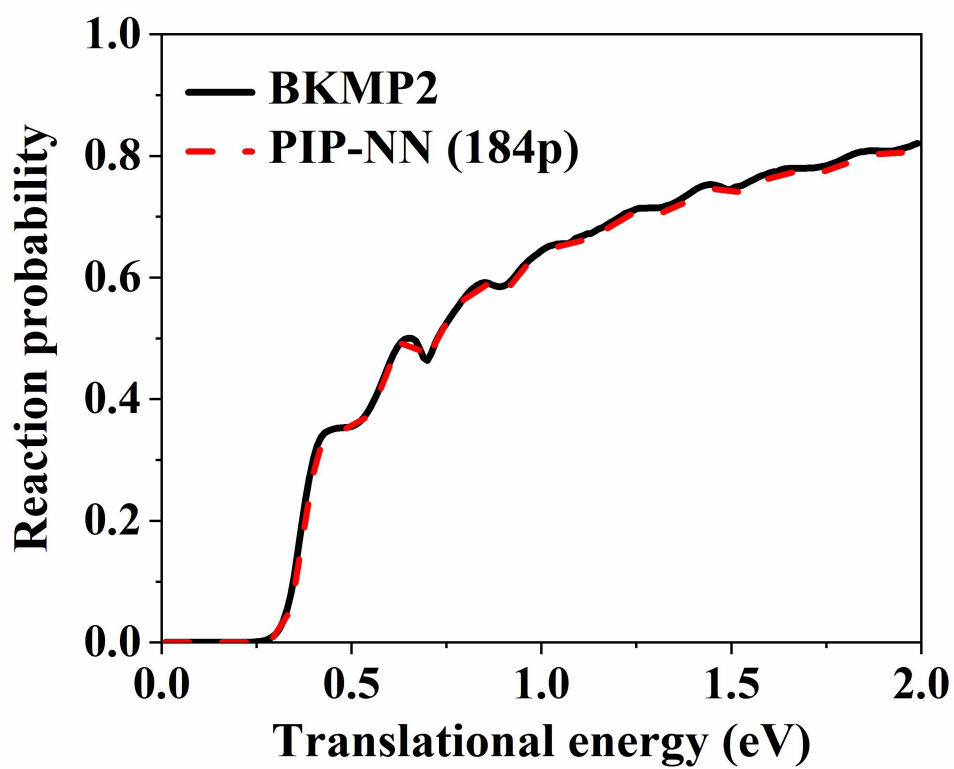

Fig. 5. Comparison of the quantum reaction probabilities obtained with the PIP-NN PES fitted to 184 points and the BKMP2 PES.

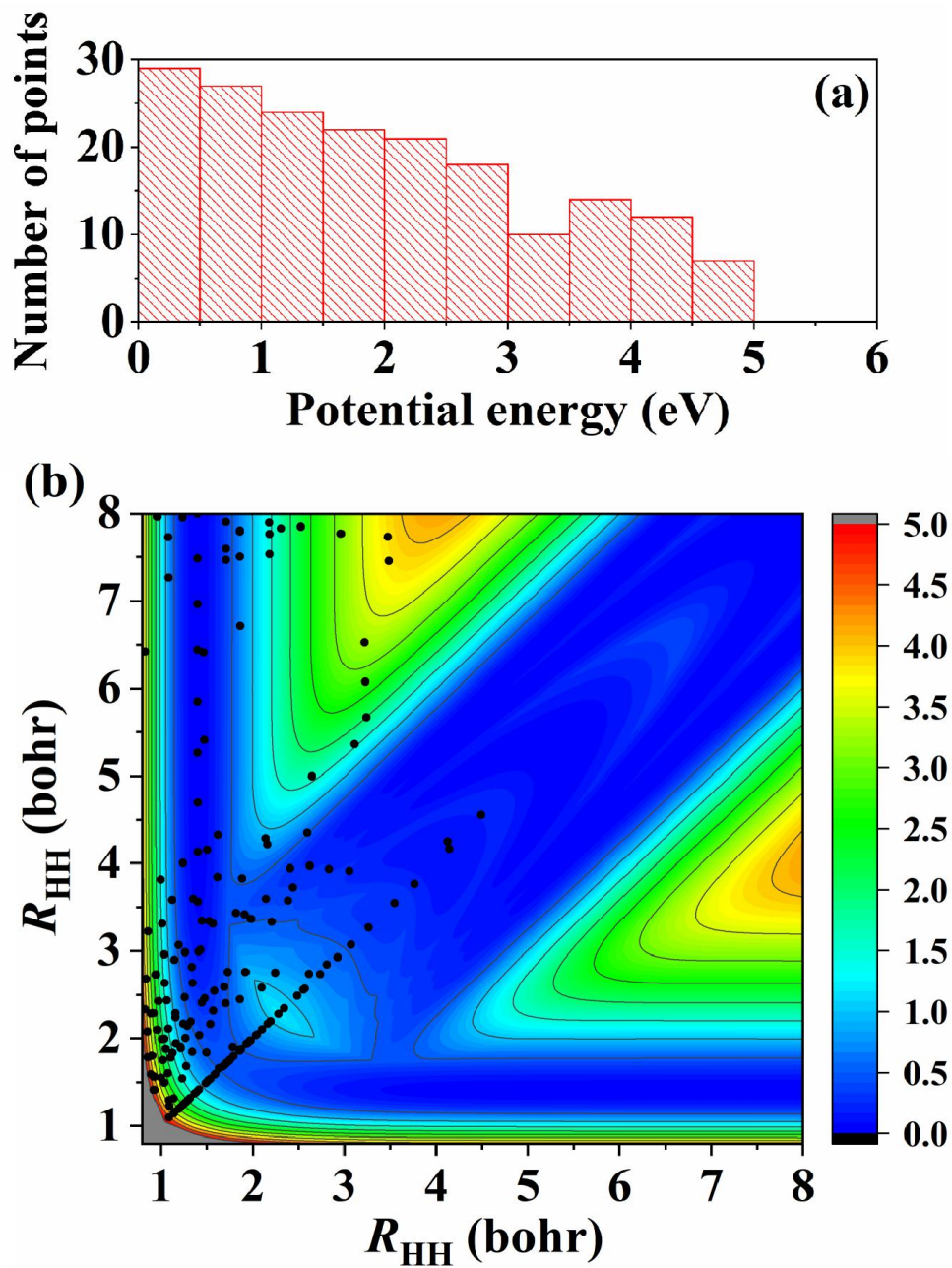

Fig. 6. Energy (a) and space (b) distributions of the 184 selected points for the H + H$_2$ reaction. The latter is projected on the potential energy (in eV) contour plot as a function of the two H-H bond lengths, with the enclosed angle $\theta_{H1H2H3}$ optimized to the minimum.

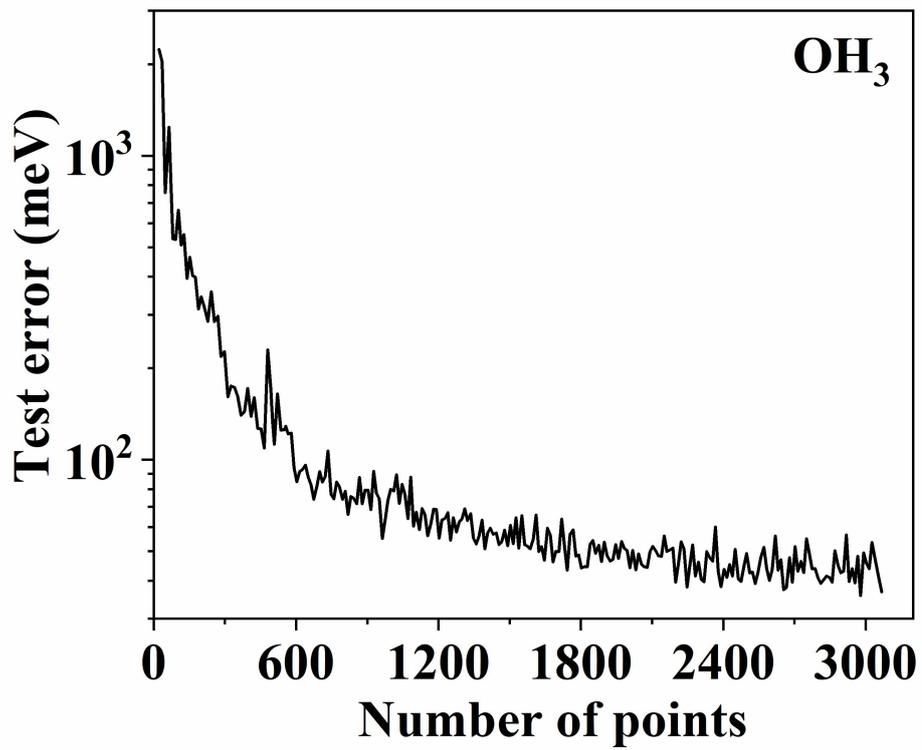

Fig. 7. Test root-mean-squared-error as a function of the number of training data points for OH$_3$ system.

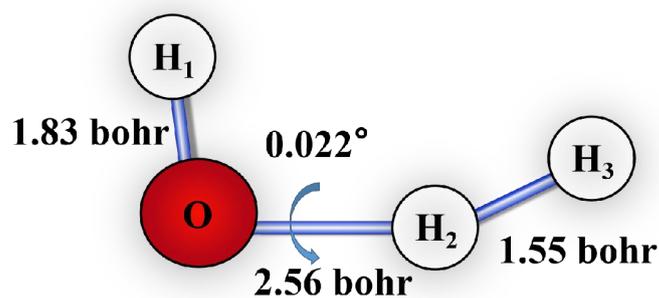

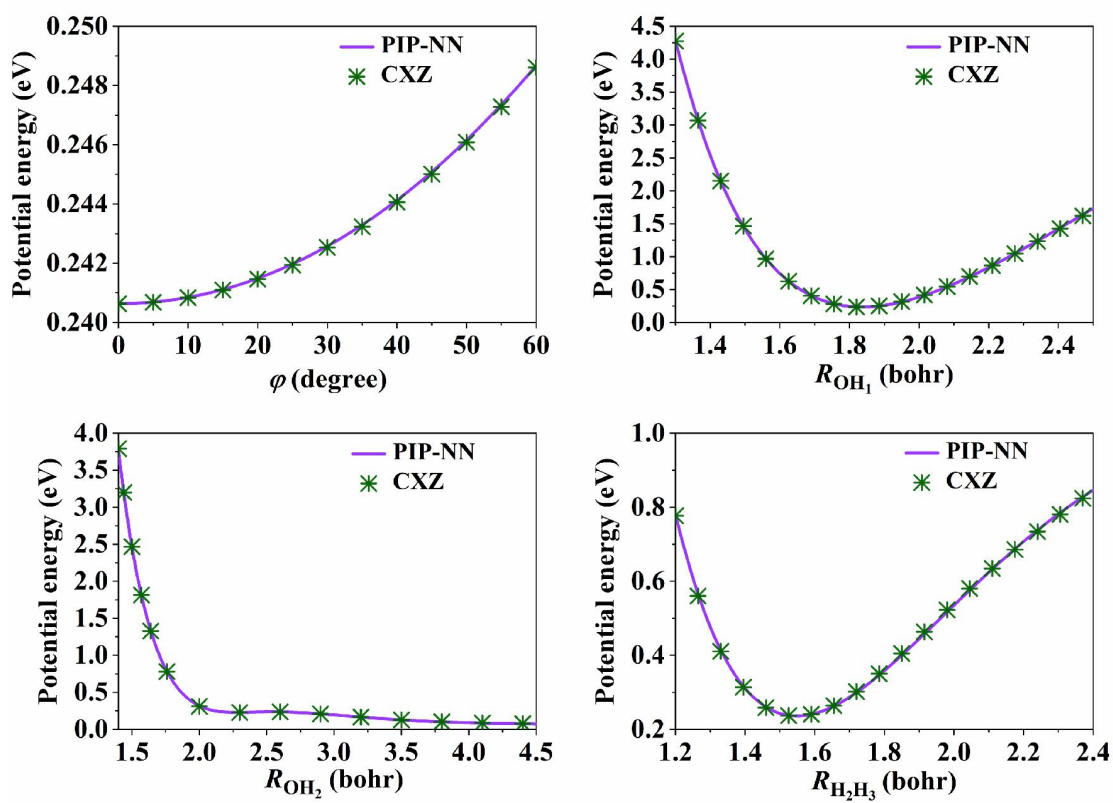

Fig. 8. Comparison of potential energy curves of the CXZ and PIP-NN PESs, as a function of internal coordinates deviating from the transition state of the abstraction reaction. $\varphi$ represents the dihedral angle between the $H_1OH_2$ and $OH_2H_3$ plane.

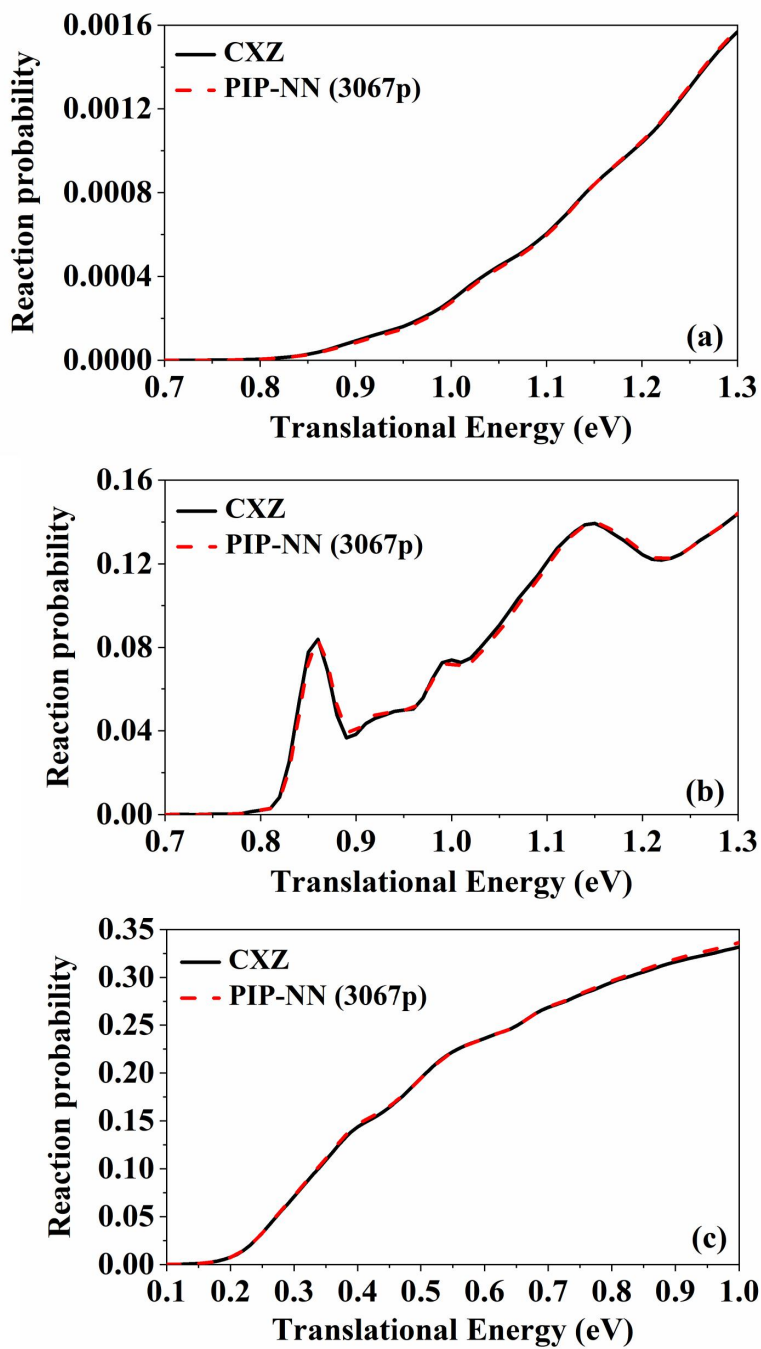

Fig. 9. Comparison of quantum reaction probabilities obtained from the PIP-NN PES fitted to 3067 points and the CXZ PES for the (a) abstraction reaction H + H$_2$O → H$_2$ + OH, (b) exchange reaction H + H′OH ⟷ HOH + H′ and (c) reverse reaction H$_2$ + OH → H + H$_2$O.

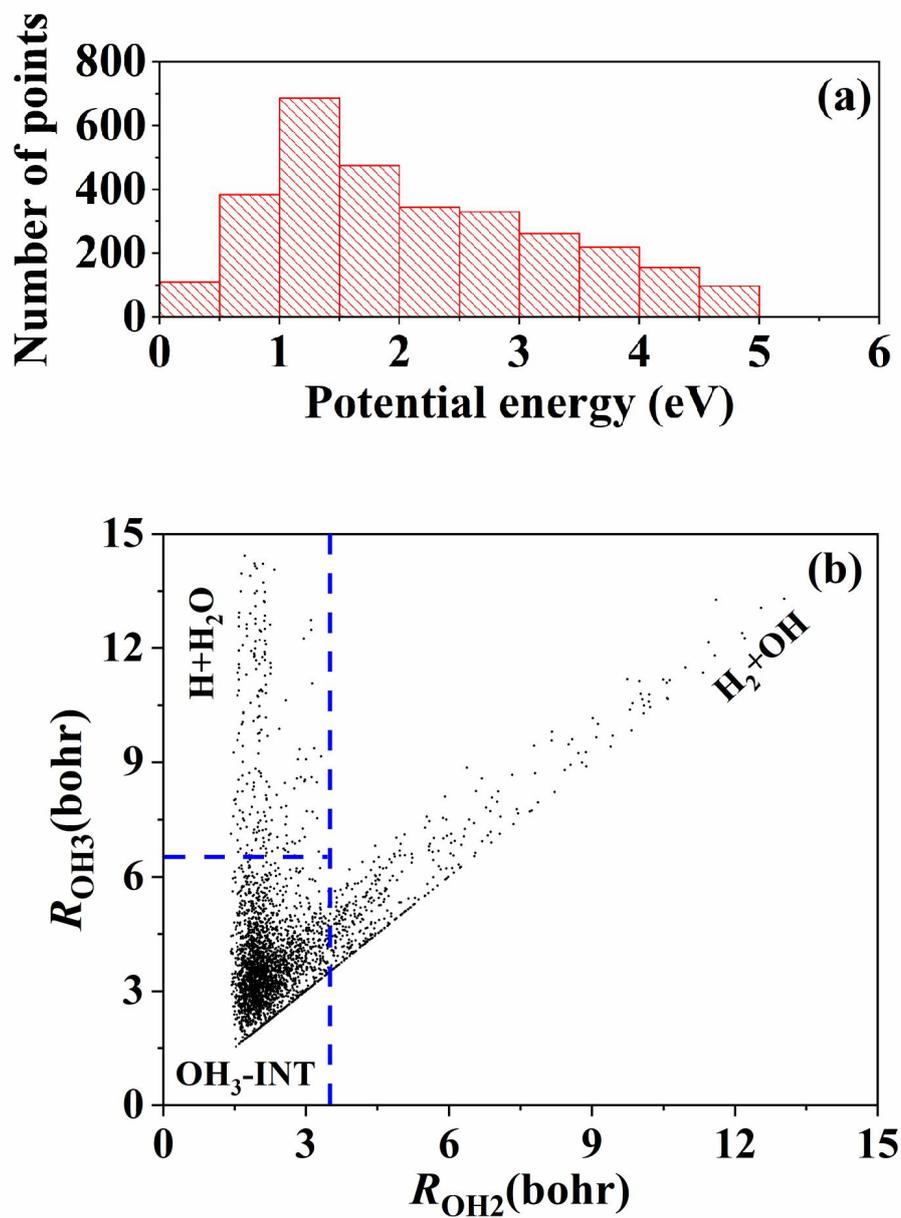

Fig. 10. Energy (a) and space (b) distributions of the 3067 points for $OH_3$ system as a function of the two longer O-H bond lengths. These three O-H bond lengths are in order of $R_{OH1} < R_{OH2} < R_{OH3}$.

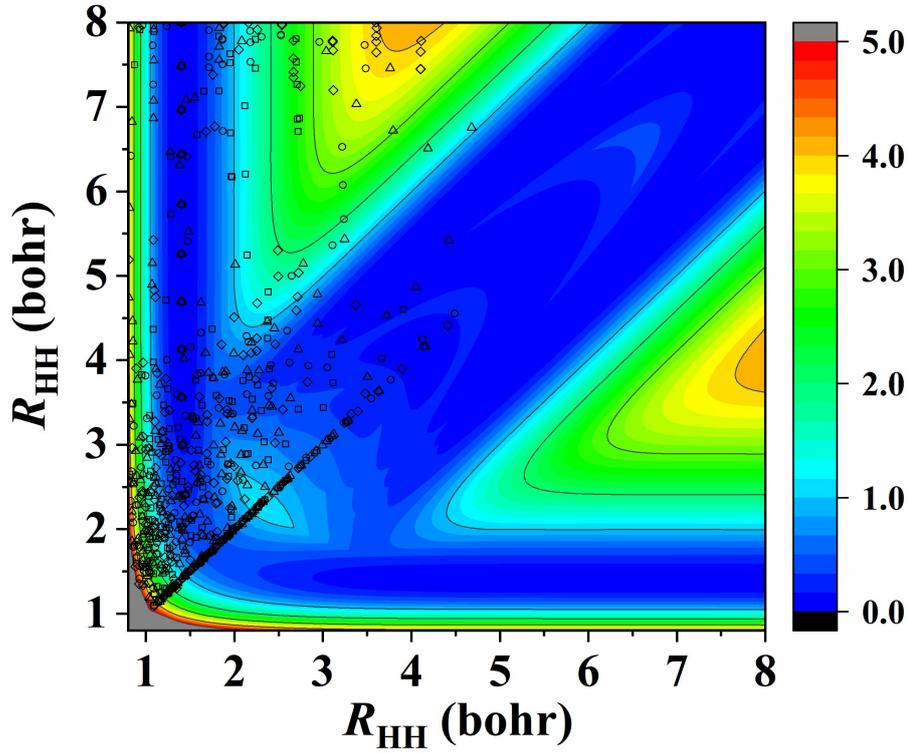

Fig. 11. Comparison of space distributions of the sampled points for the H + H$_2$ reaction projected on the potential energy (in eV) contour plot as a function of the two H-H bond lengths, with the enclosed angle $\theta_{H1H2H3}$ optimized to the minimum. Four groups of points selected with different parameters are shown as follows, (circles) $\alpha$ decays from 1.5 to 0.2 eV$^{-2}$, $\Delta\alpha$=0.1 eV$^{-2}$, 6 points per iteration, in total 184 points; (triangles) $\alpha$ decays from 1.5 to 0.2 eV$^{-2}$, $\Delta\alpha$=0.1 eV$^{-2}$, 12 points per iteration, in total 218 points; (diamonds) $\alpha$ decays from 1.5 to 0.2 eV$^{-2}$, $\Delta\alpha$=0.2 eV$^{-2}$, 6 points per iteration, in total 206 points; (squares) $\alpha$ decays from 1.5 to 0.1 eV$^{-2}$, $\Delta\alpha$=0.1 eV$^{-2}$, 6 points per iteration, in total 187 points;

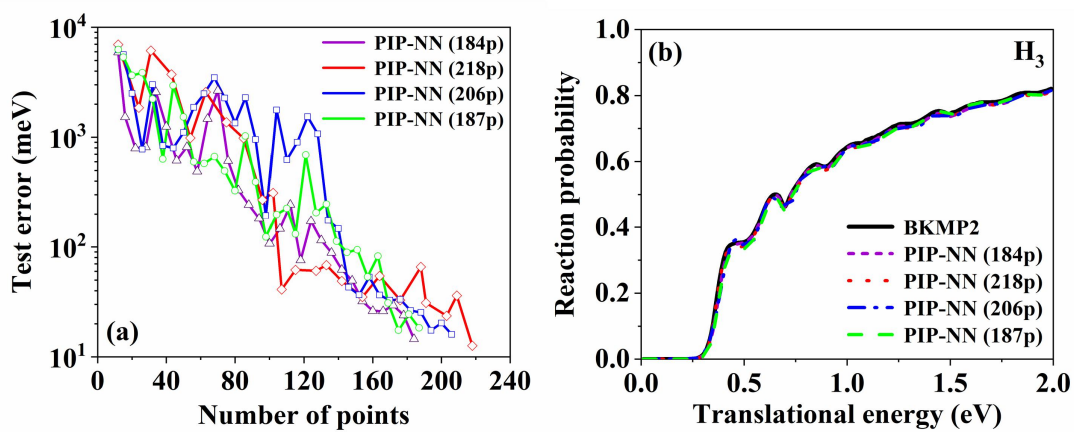

Fig. 12. (a) Comparison of the test RMSEs as a function of the number of training data points for H$_3$ system, during four different training processes with different parameters. (b) Comparison of the quantum reaction probabilities obtained on the four resulting PIP-NN PESs and of the BKMP2 PES. Four groups of points are the same as those described in Fig. 11.